\documentclass[11pt]{article}
\def\half{\mbox{\small $\frac{1}{2}$}}

\def\vec#1{\ifmmode
\mathchoice{\mbox{\boldmath$\displaystyle\bf#1$}}
{\mbox{\boldmath$\textstyle\bf#1$}}
{\mbox{\boldmath$\scriptstyle\bf#1$}}
{\mbox{\boldmath$\scriptscriptstyle\bf#1$}}\else
{\mbox{\boldmath$\bf#1$}}\fi}

\usepackage{amssymb}
\usepackage{graphicx}
\usepackage{subfigure}

\begin{document}

\begin{center}
{\Large Asymptotic distribution for two-sided tests with lower and upper boundaries on the parameter of interest.}
\end{center}

\vspace{.5 cm}

\begin{center}
Glen Cowan$^1$, Kyle Cranmer$^2$,
Eilam Gross$^3$, Ofer Vitells$^3$
\end{center}

\vspace{0.5 cm}

\noindent 
{\footnotesize 
$^1$  Physics Department, Royal Holloway, University of London, 
Egham, TW20 0EX, U.K. \\
$^2$   CCPP, Physics Department, New York University, New York, NY 10003, 
U.S.A. \\
$^3$  Weizmann Institute of Science, Rehovot 76100, Israel
}

\begin{abstract}
We present the asymptotic distribution for two-sided tests based on the profile likelihood ratio with lower and upper boundaries on the parameter of interest.  This situation is relevant for branching ratios and the elements of unitary matrices such as the CKM matrix.
\end{abstract}

%
%


\section{Introduction}

In Ref.~\cite{Cowan:2010js}, the asymptotic distribution of several test statistics based on the profile likelihood ratio are presented.  In particular, the test statistics for one- and two-sided tests of a single parameter of interest $\mu$ in the presence of nuisance parameters $\vec\theta$ are given cases where the parameter of interest is unbounded and bounded with $\mu \ge 0$.  Here we present the asymptotic distribution for two-sided tests based on the profile likelihood ratio with lower and upper boundaries on the parameter of interest.  This situation is relevant for branching ratios and the elements of unitary matrices such as the CKM matrix.  

We consider the case where $\mu \in [\mu_- , \mu_+]$.  Using the notation of Ref.~\cite{Cowan:2010js}, the maximum likelihood estimate of $\mu$ is denoted $\hat\mu$ and $\sigma^2=\textrm{var}[\hat\mu]$.  Various approaches to estimating $\sigma$ are presented in Ref.~\cite{Cowan:2010js}.  In order to use the results of Wilks and Wald, the strategy is to consider the situation when $\mu$ is unbounded and impose the boundary in the test statistic itself.  Specifically, the test statistic is
\begin{equation}
\label{eq:tmutilde} 
\tilde{t}_{\mu} = 
\left\{ \! \! \begin{array}{ll}
               - 2 \ln \frac{L(\mu, \hat{\hat{\vec{\theta}}}(\mu))}
                {L(\mu_-, \hat{\hat{\theta}}(\mu_-))}  
                & \quad \hat{\mu} \le \mu_-  \; \\*[0.2 cm]
               -2 \ln \frac{L(\mu, \hat{\hat{\vec{\theta}}}(\mu))}
                {L(\hat{\mu}, \hat{\vec{\theta}})}
&  \quad \mu_- < \hat{\mu} <  \mu_+  \; \\*[0.2 cm]
               - 2 \ln \frac{L(\mu, \hat{\hat{\vec{\theta}}}(\mu))}
                {L(\mu_+, \hat{\hat{\theta}}(\mu_+))}  
                & \quad \hat{\mu} \ge \mu_+  \;, \\*[0.2 cm]
              \end{array}
       \right.
\end{equation}
where $\hat{\hat{\vec{\theta}}}(\mu)$ is the conditional maximum likelihood estimate of $\vec\theta$ given $\mu$.

\section{Result}


The relationship between $\tilde{t}_{\mu}$ and $\hat\mu$ can be inverted to obtain $\hat\mu(\tilde{t}_{\mu})$ for the two branches $\hat{\mu} \le \mu$ and $\hat{\mu} > \mu$.  Wald's theorem states that 
\begin{equation}
f(\hat{\mu}|\mu) \approx  \frac{1}{\sqrt{2 \pi} \sigma} \exp\left( - \frac{(\hat\mu-\mu')^2}{2\sigma^2} \right)\; .
\end{equation}
Through a straightforward change of variables, one can obtain the distribution for the test statistic $\tilde{t}_{\mu}$.  The pdf $f(\tilde{t}_{\mu} | \mu^{\prime})$
is found to be
\begin{eqnarray}
\label{eq:ftildeqmmp} 
f(\tilde{t}_{\mu}|\mu^{\prime}) & = & f_L(\tilde{t}_{\mu}|\mu^{\prime})+f_R(\tilde{t}_{\mu}|\mu^{\prime})
\end{eqnarray}
with
\begin{eqnarray}
\label{eq:ftildeqmmp} 
f_L(\tilde{t}_{\mu}|\mu^{\prime}) & = & 
 \: \left\{ \! \! \begin{array}{lll}
\frac{1}{2} \frac{1}{\sqrt{2 \pi}} \frac{1}{\sqrt{\tilde{t}_{\mu}}}
\exp \left[ -\frac{1}{2} \left( \sqrt{\tilde{t}_{\mu}} -
\frac{\mu - \mu^{\prime}}{\sigma} \right)^2 \right]
                 & \quad \tilde{t}_{\mu} \le \delta_-^2  \\*[0.5 cm]
\frac{1}{\sqrt{2 \pi}} \frac{1}{2\delta_-} \exp \left[
-\frac{1}{2} \frac{ (\tilde{t}_{\mu} - 
(\delta_-^2 - 2 \delta_- \delta_-^{\prime}) )^2 }
{(2 \delta_-)^2} \right] 
                 &  \quad \tilde{t}_{\mu} > \delta_-^2 
              \end{array}
       \right.
\end{eqnarray}
and
\begin{eqnarray}
\label{eq:ftildeqmmp} 
f_R(\tilde{t}_{\mu}|\mu^{\prime}) & = & 
 \: \left\{ \! \! \begin{array}{lll}
\frac{1}{2} \frac{1}{\sqrt{2 \pi}} \frac{1}{\sqrt{\tilde{t}_{\mu}}}
\exp \left[ -\frac{1}{2} \left( \sqrt{\tilde{t}_{\mu}} +
\frac{\mu - \mu^{\prime}}{\sigma} \right)^2 \right]
                 &  \quad \tilde{t}_{\mu} \le \delta_+^2  \\*[0.5 cm]
\frac{1}{\sqrt{2 \pi}}\frac{1}{2\delta_+} \exp \left[
-\frac{1}{2} \frac{ (\tilde{t}_{\mu} + 
(\delta_+^2 - 2 \delta_+ \delta_+^{\prime}) )^2 }
{(2 \delta_+)^2} \right] 
                 &  \quad \tilde{t}_{\mu} > \delta_+^2 \;,
              \end{array}
       \right.
\end{eqnarray}
where the dimensionless variables $\delta_-=(\mu-\mu_-)/\sigma$, $\delta_-^\prime=(\mu^\prime-\mu_-)/\sigma$ , $\delta_+ = (\mu - \mu_+)/\sigma$, and $\delta^\prime_+ = (\mu^\prime - \mu_+)/\sigma$ are used to simplify the expressions.

 \vspace{1em}
 The special case $\mu = \mu^{\prime}$ is therefore
\begin{equation}
\label{eq:ftildetmm} 
f(\tilde{t}_{\mu}|\mu) = 
 \: \left\{ \! \! 
 \begin{array}{ll}
\frac{1}{\sqrt{2 \pi}}
\frac{1}{\sqrt{\tilde{t}_{\mu}}} e^{- \tilde{t}_{\mu} / 2 } 
& \quad \tilde{t}_{\mu} \le \delta_{C}^2  \; \\*[0.5 cm]
\frac{1}{2} \frac{1}{\sqrt{2 \pi}}
\frac{1}{\sqrt{\tilde{t}_{\mu}}} e^{- \tilde{t}_{\mu} / 2 } \: + \:
\frac{1}{\sqrt{2 \pi}} \frac{1}{2 \delta_{C}} \exp \left[ - \half 
\frac{ (\tilde{t}_{\mu} + \delta_{C}^2)^2}{ (2 \delta_{C})^2 }  \right] 
& \delta_{C}^2 <  \tilde{t}_{\mu} < \delta_{F}^2 \\*[0.5 cm]
 \frac{1}{\sqrt{2 \pi}} \frac{1}{ 2 \delta_{C}} \exp \left[ - \half 
\frac{ (\tilde{t}_{\mu} + \delta_{C}^2)^2}{ (2 \delta_{C})^2 }  \right] 
\: + \:
\frac{1}{\sqrt{2 \pi} } \frac{1}{2 \delta_{F}} \exp \left[ - \half 
\frac{ (\tilde{t}_{\mu} + \delta_{F}^2)^2}{ (2 \delta_{F})^2 }  \right] 
& \quad \tilde{t}_{\mu} \ge \delta_{F}^2 \;, 
 \end{array}
 \right.
\end{equation}
where $\delta_{C} = \min[\mu-\mu_-, \mu_+-\mu ]/\sigma$ and $\delta_{F} = \max[\mu-\mu_-, \mu_+-\mu]/\sigma$.

\noindent The corresponding cumulative distribution is

\begin{equation}
\label{eq:tildetmmpcdf} 
F(\tilde{t}_{\mu}|\mu^{\prime}) = F_L(\tilde{t}_{\mu}|\mu^{\prime}) + F_R(\tilde{t}_{\mu}|\mu^{\prime})
\end{equation}
with
\begin{eqnarray}
\label{eq:ftildeqmmp} 
F_L(\tilde{t}_{\mu}|\mu^{\prime}) & = & 
 \: \left\{ \! \! \begin{array}{lll}
\Phi \left[  \left( \sqrt{\tilde{t}_{\mu}} -
\frac{\mu - \mu^{\prime}}{\sigma} \right) \right]-\half
               \quad  & \tilde{t}_{\mu} \le \delta_-^2  \\*[0.5 cm]
\Phi \left[ \frac{ \tilde{t}_{\mu} - 
(\delta_-^2 - 2 \delta_- \delta_-^{\prime}) }
{2 \delta_-} \right] -\half
                 &   \tilde{t}_{\mu} > \delta_-^2 
              \end{array}
       \right.
\;
\end{eqnarray}
and
\begin{eqnarray}
\label{eq:ftildeqmmp} 
F_R(\tilde{t}_{\mu}|\mu^{\prime}) & = & 
 \: \left\{ \! \! \begin{array}{lll}
\Phi \left[  \left( \sqrt{\tilde{t}_{\mu}} +
\frac{\mu - \mu^{\prime}}{\sigma} \right) \right] - \half
                 \quad & \tilde{t}_{\mu} \le \delta_+^2  \\*[0.5 cm]
\Phi \left[
 \frac{ \tilde{t}_{\mu} + 
(\delta_+^2 - 2 \delta_+ \delta_+^{\prime})  }
{2 \delta_+} \right] -\half
                 &   \tilde{t}_{\mu} > \delta_+^2 \;,
              \end{array}
       \right.
\;
\end{eqnarray}
where $\Phi(x)$ is the cumulative probability distribution of the standard normal distribution.

The special case $\mu = \mu^{\prime}$ is therefore
\begin{equation}
\label{eq:ftildetmm} 
F(\tilde{t}_{\mu}|\mu) = 
 \: \left\{ \! \! 
 \begin{array}{ll}
2\Phi\left[\sqrt{ \tilde{t}_{\mu}} \right]-1
& \quad \tilde{t}_{\mu} \le \delta_{C}^2  \; \\*[0.5 cm]
\Phi\left[\sqrt{ \tilde{t}_{\mu}} \right] \: + \:
\Phi \left[  
\frac{ \tilde{t}_{\mu} \: + \:\delta_{C}^2}{ 2 \delta_{C} }  \right] -1
\quad & \delta_{C}^2 <  \tilde{t}_{\mu} < \delta_{C}^2 \\*[0.5 cm]
\Phi \left[ 
\frac{ \tilde{t}_{\mu} + \delta_{C}^2}{ 2 \delta_{F} }  \right] 
\: + \:
\Phi \left[ 
\frac{ \tilde{t}_{\mu} + \delta_{F}^2}{ 2 \delta_{F} }  \right] -1
& \quad \tilde{t}_{\mu} > \delta_{F}^2 \;.
 \end{array}
 \right.
\;
\end{equation}

\section{The critical cutoff}

Confidence intervals are defined as the set of $\mu$ where the test statistic $\tilde{t}_\mu$ is less than or equal to a critical cutoff $k_\alpha(\mu)$.  The cutoff is chosen to insure the desired coverage probability.  For a $100 (1-\alpha)$ \% confidence level interval, the cutoff is defined by $F(k_\alpha(\mu)|\mu) = 1-\alpha$.

When there are no boundaries, the distribution of the test statistic follows a $\chi^2$ distribution and the cutoff is constant.  Specifically, it is given by $(\Phi^{-1}(1-\alpha/2))^2$, which gives the familiar values of 3.84 for 95\%, 2.71 for 90\%, and 1 for 68\% confidence intervals.

When will the boundary matter?  The critical cutoff is modified for $\mu<\mu_- + \sigma \Phi^{-1}(1-\alpha/2)$ and $\mu>\mu_+ -\sigma \Phi^{-1}(1-\alpha/2)$.  Thus edges of a confidence interval using the standard cutoff are fine if they fall in the intermediate $\mu$ range; however, they will over-cover if they are near the boundaries.  As $\sigma$ increases the range of $\mu$ with a modified cutoff grows.  Once $\sigma>\sigma_{\rm crit} = (\mu_+ - \mu_-)/(2\sqrt{t_\alpha})$, then there is no region of $\mu$ where the cutoff is not affected.

When testing at the boundary, the critical value $k_\alpha(\mu_-) = k_\alpha(\mu_+)$ is always affected.  For large values of $\delta = (\mu_+-\mu_-)/\sigma$ (ie. when $\hat\mu$ is well measured with respect to the range of $\mu$) only one boundary is important; however, for small values of $\delta$ (ie. when $\hat\mu$ is poorly measured with respect to the range of $\mu$) both boundaries are important.  The cutoff at the boundary is given by
\begin{equation}
k_\alpha(\mu_-) = k_\alpha(\mu_+) =  \: \left\{ \! \! 
 \begin{array}{ll}
2\Phi^{-1}(1-\alpha)) \delta - \delta^2  \quad& \delta \le \Phi^{-1}(1-\alpha) \\*[0.5 cm]
(\Phi^{-1}(1-\alpha))^2   & \delta > \Phi^{-1}(1-\alpha) 
 \end{array}
 \right.
\end{equation}
Note that for a 95\% confidence interval, if $\sigma >  (\mu_+-\mu_-)/1.64$, then the far away boundary reduces the critical cutoff below the 2.71 one might expect from the presence of the boundary being tested and it is significantly smaller than the 3.84 cutoff one has from assuming a $\chi^2$ distribution neglecting any boundary effects.

\begin{figure}[htb]
\begin{center}
{\includegraphics[width=.7\textwidth]{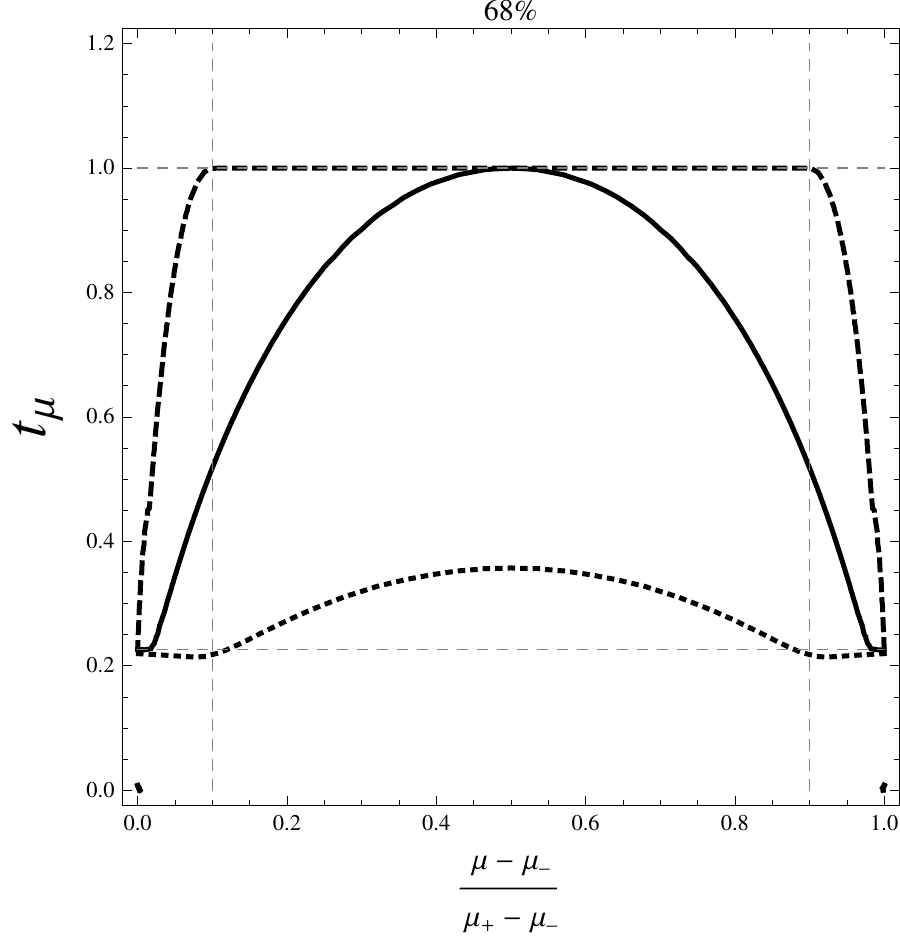}}
\caption{The critical cutoff for 68\% confidence level interval for several cases.
The upper horizontal line corresponds to the naive cutoff neglecting any boundaries of $(\Phi^{-1}(1-\alpha/2))^2$. The lower horizontal line corresponds to the cutoff of $(\Phi^{-1}(1-\alpha))^2$, which is appropriate when testing on the boundary when $\sigma$ is sufficiently small.  
The dashed curve corresponds to $\sigma=(\mu_+ - \mu_-)/10$ where the cutoff is only affected for 
$\mu<\mu_- + \sigma \Phi^{-1}(1-\alpha/2)$ and $\mu> \mu_+ -\sigma \Phi^{-1}(1-\alpha/2)$ (vertical lines).   The solid curve corresponds to the cutoff for $\sigma=\sigma_{\rm crit}$; for any value of $\sigma>\sigma_{\rm crit}$, the cutoff is affected for all values of $\mu$.  Finally, the dotted curve shows the case of $\sigma=1.2 (\mu_+-\mu_-)/ \Phi^{-1}(1-\alpha)$, where $\sigma>(\mu_+-\mu_-)/ \Phi^{-1}(1-\alpha)$ means that the critical cutoff on the boundaries is affected by the faraway boundary.    }
\label{fig:cutoff68}
\end{center}
\end{figure}

\begin{figure}[htb]
\begin{center}
{\includegraphics[width=.5\textwidth]{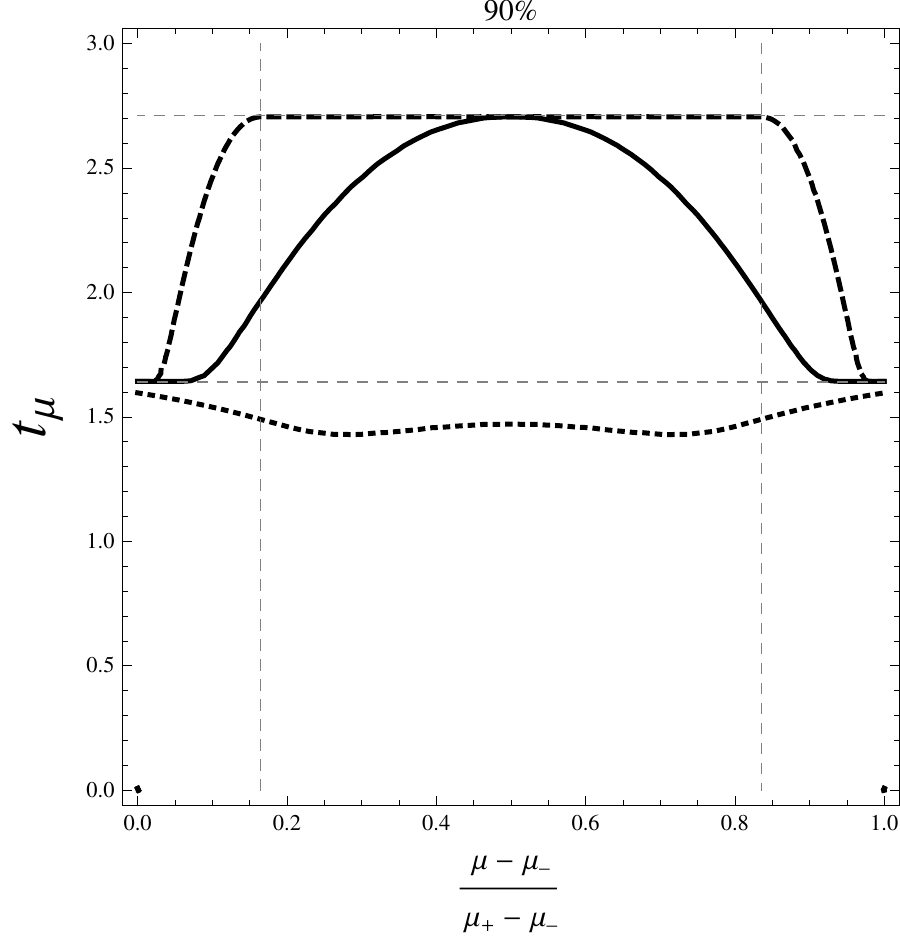}}
\caption{The same as Figure~\ref{fig:cutoff68} for 90\%.}
\label{fig:cutoff90}
\end{center}
\end{figure}

\begin{figure}[htb]
\begin{center}
{\includegraphics[width=.5\textwidth]{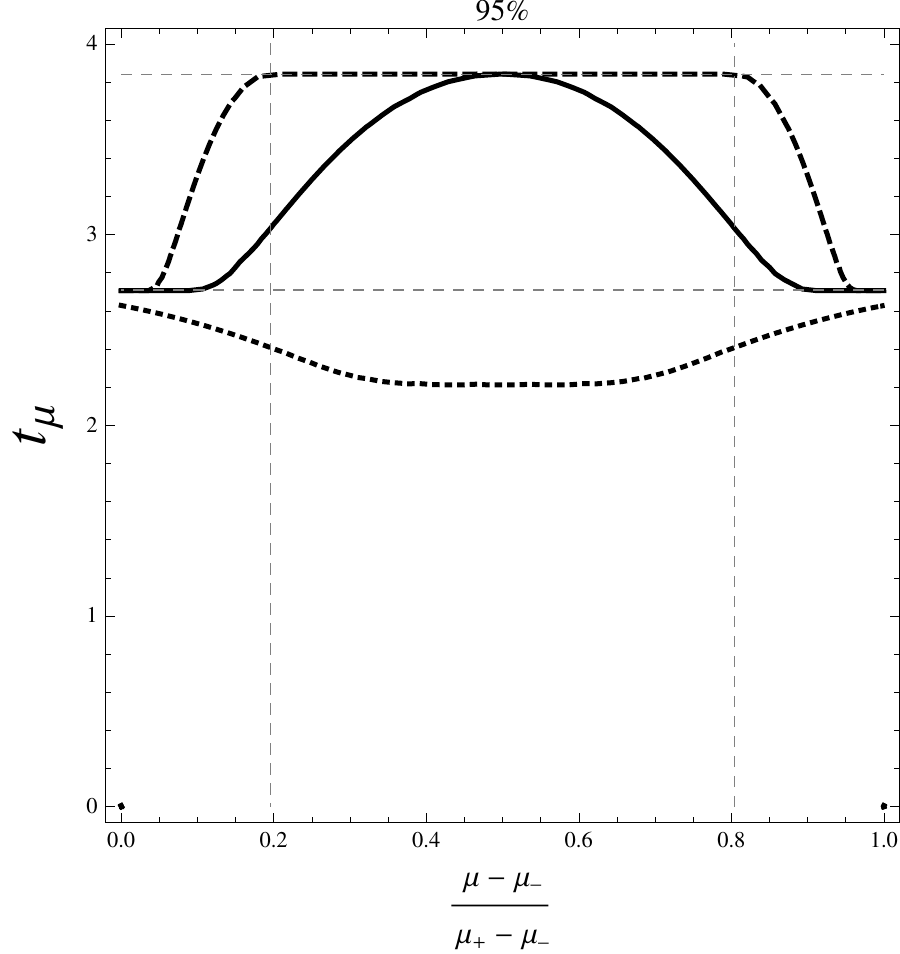}}
\caption{The same as Figure~\ref{fig:cutoff68} for 95\%.}
\label{fig:cutoff95}
\end{center}
\end{figure}

Figures~\ref{fig:cutoff68}-\ref{fig:cutoff95} show the critical cutoff $k_\alpha(\mu)$ for 68\%, 90\%, and 95\% confidence intervals for $\sigma \ll \sigma_{\rm crit}$, $\sigma = \sigma_{\rm crit}$, and $\sigma \sim (\mu_+-\mu_-)/\Phi^{-1}(1-\alpha)$.

\section{Conclusions}

The presence of both lower- and upper-boundaries on a parameter of interest is a common situation in particle physics.  For example, branching ratios and elements of unitary matrices are bounded between 0 and 1.  The formulae presented here are essentially the asymptotic versions of the Feldman-Cousins approach~\cite{Feldman:1997qc} extended to incorporate nuisance parameters via the use of the profile likelihood ratio test statistic as in Ref.~\cite{Cowan:2010js}.



\begin{thebibliography}{9}
\bibitem{Cowan:2010js} 
  G.~Cowan, K.~Cranmer, E.~Gross and O.~Vitells,
  Eur.\ Phys.\ J.\ C {\bf 71}, 1554 (2011)
  [arXiv:1007.1727 [physics.data-an]].
  
\bibitem{Feldman:1997qc} 
  G.~J.~Feldman and R.~D.~Cousins,
  Phys.\ Rev.\ D {\bf 57}, 3873 (1998)
  [physics/9711021 [physics.data-an]].
  
\end{thebibliography}
\end{document}